\documentclass[aps,pre,superscriptaddress,showpacs,nofootinbib,floatfix,reprint]{revtex4-1}
\usepackage{amsmath}
\usepackage{amssymb}
\usepackage{graphicx}
\usepackage{float}
\usepackage{subfigure}
\usepackage{color}
\usepackage[utf8x]{inputenc}

\begin{document}
\title{Title}
\author{Nathann T. Rodrigues}
\email{nathan.rodrigues@ufv.br}
\author{Tiago J. Oliveira}
\email{tiago@ufv.br}
\affiliation{Departamento de Física, Universidade Federal de Viçosa, 36570-900, Vi\c cosa, MG, Brazil}
\date{\today}

\title{Thermodynamic behavior of binary mixtures of hard spheres: Semianalytical solutions on a Husimi lattice built with cubes}

\begin{abstract}
We study binary mixtures of hard particles, which exclude up to their $k$th nearest neighbors ($k$NN) on the simple cubic lattice and have activities $z_k$. In the first model analyzed, point particles (0NN) are mixed with 1NN ones. The grand-canonical solution of this model on a Husimi lattice built with cubes unveils a phase diagram with a fluid and a solid phase separated by a continuous and a discontinuous transition line which meet at a tricritical point. A density anomaly, characterized by minima in isobaric curves of the total density of particles against $z_0$ (or $z_1$), is also observed in this system. Overall, this scenario is identical to the one previously found for this model when defined on the square lattice. The second model investigated consists of the mixture of 1NN particles with 2NN ones. In this case, a very rich phase behavior is found in its Husimi lattice solution, with two solid phases - one associated with the ordering of 1NN particles ($S1$) and the other with the ordering of 2NN ones ($S2$) -, beyond the fluid ($F$) phase. While the transitions between $F$-$S2$ and $S1$-$S2$ phases are always discontinuous, the $F$-$S1$ transition is continuous (discontinuous) for small (large) $z_2$. The critical and coexistence $F$-$S1$ lines meet at a tricritical point. Moreover, the coexistence $F$-$S1$, $F$-$S2$ and $S1$-$S2$ lines meet at a triple point. Density anomalies are absent in this case.
\end{abstract}

\maketitle

\section{Introduction}
\label{secIntro}

Hard spheres (and hard disks in two-dimensions) interacting only through an infinite hard-core potential form the simplest interacting models for a classical fluid. Such athermal models have received a lot of attention in the last decades, because they allow us to investigate only the entropic effect on the thermodynamics of fluids and for their appealing role in the modeling of granular systems \cite{Applications}. In the monodisperse case, a first-order fluid-solid transition has been observed in numerical simulations of hard spheres \cite{Wood,Alder,Hoover}. For binary mixtures of particles with dissimilar sizes, several solid phases may arise [see, e.g, \cite{Eldridge} and refs. therein]. Moreover, beyond the freezing transition, the existence of solid-solid and fluid-fluid demixing transitions are also expected. For nonadditive mixtures of hard particles a demixing is expected because particles fill the space more effectively when separated into pure phases \cite{Widom,Melnyk,Santos,Louis}. Numerical simulations of binary nonadditive mixtures of hard spheres have indeed confirmed this \cite{Louis,Dijkstra}. In the case of additive mixtures the scenario is a bit more controversial, with the possibility of fluid-fluid phase separation being demonstrated in some recent analytical treatments, provided that the particle's sizes are dissimilar enough \cite{BH91,Lekkerkerker932}, in contrast with elder analytical approaches ruling out the demixing \cite{LR64}. However, in general, fluid-solid or solid-solid coexistence can preempt the fluid-fluid transition [see, e.g, \cite{LafuenteCuesta,LafuenteCuesta2} and refs. therein].

When defined on a lattice, the hard spheres (and disks) are approximated by $k$NN particles, i.e, particles which forbid up to their $k$th nearest neighbor (NN) sites of being occupied. The $0$NN case consists of point particles which do not interact and, so, do not undergo any transition. On the other hand, pure $k$NN models with $k\geqslant 1$ are known to present fluid-solid transitions. For instance, the 1NN case on the triangular lattice is the famous hard-hexagon model exactly solved by Baxter \cite{baxterHH,baxterBook}, whose continuous fluid-solid transition belongs to a universality class different from the Ising one found for this model on the square lattice \cite{GuoBlote}. In the cubic lattice, 3D Ising exponents have been found for the 1NN model \cite{Yamagata,HB}. While recent works indicate that the transition in the 2NN model on the square lattice is continuous, its universality class is still a subject of debate (see e.g. \cite{Heitor,Blote2NN} and references therein). On the cubic lattice, a discontinuous fluid-solid transition has been found for this model both on numerical simulations \cite{Panagiotopoulos} and mean-field approximations \cite{Nathann19}. For larger $k$'s, discussions on the behavior can be found in Refs. \cite{Heitor,Rajesh} for the square and \cite{Panagiotopoulos} for the cubic lattice. 

As an aside, let us notice that entropy-driven transitions have been also investigated for other particle shapes, such as cubes \cite{Rajeshcubes}, dimers \cite{dimers}, rectangles \cite{rectangles}, rods \cite{rods}, triangles \cite{Nienhuistri}, Y-shaped \cite{RajeshY}, etc., on different lattices. Some binary mixtures of hard lattice gases with isotropic \cite{Dijkstra,Schmidt,Brader,frenkel0nn1nn,frenkelcubic1,frenkelcubic2,Dickman95} and anisotropic particles \cite{Roiji,Wensink,Dubois,Varga,Mederos,Schmidt02,Heras,Jurgenrods} have been also analyzed, which in several cases exhibit fluid-fluid or solid-solid  demixing transitions.

Curiously, however, mixtures of $k$NN particles are far less explored. For instance, the 0NN-1NN case has been investigated on the square lattice by means of series expansions \cite{poland} and transfer matrix calculations \cite{Jim01,tiago15}. These later studies revealed a grand-canonical phase diagram featured by a fluid and a solid phase separated by a continuous and a discontinuous transition line, which meet at a tricritical point. The same scenario was found in a mean-field solution of this model on the Bethe lattice \cite{tiago11}. Interestingly, such model display a thermodynamic anomaly characterized by minima in isobaric curves of the total density of particles \cite{tiago15,tiago11}. On the triangular lattice, some authors \cite{Lekkerkerker93,Lekkerkerker95} claimed to have numerically found a fluid-fluid transition, yielding three stable (gas-liquid-solid) phases for the 0NN-1NN mixture. However, evidence against the fluid-fluid demixing have been presented in other works \cite{frenkel0nn1nn,Nienhuis}. Particularly, in Ref. \cite{Nienhuis} strong numerical evidence for a phase diagram similar to the one just discussed for the square lattice was presented. 

On the cubic lattice, for the best of our knowledge, only the 0NN-2NN mixture has been considered up to now, via a semi-analytical (mean-field) solution of the model on a Husimi lattice built with cubes \cite{Nathann19}. Interestingly, a stable fluid-fluid transition was found in this system, so that three stable phases are present in its grand-canonical phase diagram: two fluids (being one regular and other featured by a dominance of point particles) and a solid phase. These phases are separated by first-order transition lines which meet at a triple point, whilst the fluid-fluid coexistence line ends at a critical point. A density anomaly similar to the one found on the square lattice 0NN-1NN model is also present in this system. In face of the scarcity of binary systems presenting fluid-fluid transitions, it turns out very important to extend the existing studies for other mixtures of $k$NN particles, in order to determine the conditions (e.g, the difference in particle size) necessary for its onset. Moreover, binary mixtures of hard particles are certainly more appealing for the modeling of real fluids or granular matter than the monocomponent case - for instance, binary mixtures of colloidal hard-spheres have been widely studied \cite{binarycolloids} -, which also justify further investigations of these systems.

\begin{figure*}[t]
\includegraphics[width=14.0cm]{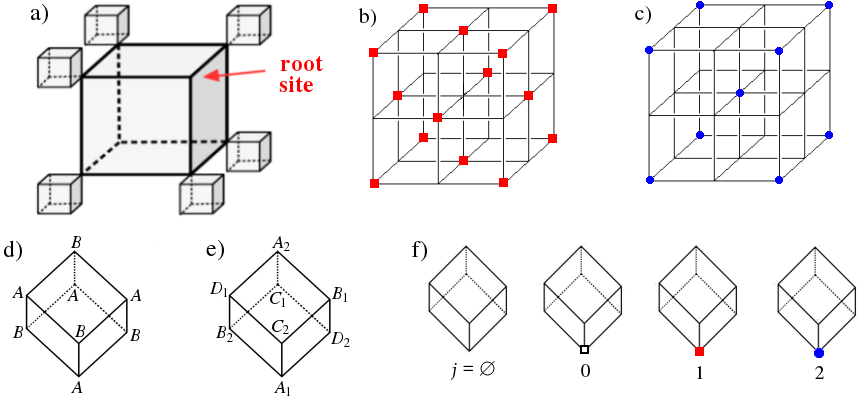}
\caption{(a) Illustration of part of a Husimi lattice built with cubes. The ground states of the solid $S1$ (b) and $S2$ (c) phases on the cubic lattice. Definitions of sublattices in an elementary cube used to solve the 0NN-1NN (d) and 1NN-2NN (e) mixtures. (f) Possible states of the root site, where 0NN, 1NN and 2NN particles are indicated by an open square, a full square and a circle, respectively.}
\label{fig1}
\end{figure*}

Here, we address this by analyzing the mixtures of 0NN-1NN and 1NN-2NN particles defined on the simple cubic lattice. From semi-analytical grand-canonical solutions of these models on Husimi lattices built with cubes (see Fig. \ref{fig1}a), we demonstrate that no fluid-fluid demixing occurs in such mixtures, in contrast with the 0NN-2NN one \cite{Nathann19}. In fact, the thermodynamic behavior of the 0NN-1NN model is qualitatively the same described above for its counterpart on the square lattice. For the 1NN-2NN mixture, a solid-solid demixing is found, between the ordered phases characteristic of the 1NN and 2NN particles, while a single fluid phase is present in the system. These three phases give rise to a very rich phase diagram, with continuous and discontinuous transitions lines, a tricritical and a triple point.

The rest of the paper is organized as follows. In Sec. \ref{secModel} we define the models and devise their solutions on the Husimi lattice built with cubes. The thermodynamic properties of the 0NN-1NN and 1NN-2NN mixtures are presented in Secs. \ref{secRes0NN1NN} and \ref{secRes1NN2NN}, respectively. In Sec. \ref{secConc} our final discussions and conclusions are summarized. Some details on the model solutions are presented in the appendix.

\section{Models and their solution on a Husimi lattice built with cubes}
\label{secModel}

We consider lattice gases consisting of binary mixtures of hard particles placed on (and centered at) the vertices of a cubic lattice. Assuming that the lattice spacing is $a$, the small particles (0NN) are cubes of lateral size $\lambda=a$, which occupy a single lattice site and do not exclude their neighbors. On the other hand, the larger 1NN (2NN) particles are cubes of lateral size $\lambda = \sqrt{2}a$ ($\lambda = \sqrt{3}a$) placed on the lattice in a way that they exclude their first (first and second) nearest neighbors. An activity $z_k$ is associated with each $k$NN particle. The mixture 0NN-2NN ($z_1=0$ case) was already investigated by us in \cite{Nathann19}, so here we will focus on the cases $z_2=0$ and $z_0=0$, corresponding respectively to the mixtures 0NN-1NN and 1NN-2NN. 

Let us remark that while the pure 0NN system ($z_1=z_2=0$) can be trivially solved and does not display any transition, the pure 1NN ($z_0=z_2=0$) and 2NN ($z_0=z_1=0$) models on the cubic lattice are known to undergo a continuous and a discontinuous transition, respectively, from disordered fluid phases to ordered solid ones \cite{Panagiotopoulos}. As illustrated in Figs. \ref{fig1}b and \ref{fig1}c, both solid phases are characterized by a sublattice ordering. In the 1NN solid ($S1$) phase one of two sublattices is preferentially occupied and in the ground state (the full occupancy limit) the density per site of 1NN particles is $\rho_1=1/2$. In the 2NN solid ($S2$) phase one of four sublattices is more occupied and the maximum density of 2NN particles is $\rho_2=1/4$. Therefore, to investigate the 0NN-1NN  mixture one has to divide the lattice into two sublattices ($A$ and $B$), as shows Fig. \ref{fig1}d. For the 1NN-2NN system, eight sublattices (as defined in Fig. \ref{fig1}e) are need in order to capture the symmetries of both $S1$ and $S2$ phases.

Following our previous study of the 0NN-2NN mixture \cite{Nathann19}, instead of investigating the models on the simple cubic lattice, we will solve them on a Husimi lattice built with cubes (see Fig. \ref{fig1}a). Let us note that the solution of a given model in the core of a Cayley tree of coordination $q$ usually corresponds to the (mean-field) Bethe approximation for this model on a regular lattice with the same coordination, and for this reason this is known as the \textit{Bethe lattice} \cite{baxterBook}. Once the Cayley tree (and so the Bethe lattice) has no loops, an improved approximation can be obtained by building such tree with polygons or polyhedrons. The core of this cactus is the \textit{Husimi lattice} \cite{Husimi}. We remark that solutions on these lattices usually provide the qualitatively correct thermodynamic behavior of the investigated models, as is indeed the case for the 0NN-1NN mixture on the square \cite{Jim01,tiago15} and Bethe lattices \cite{tiago11}. Moreover, for certain lattice gases even a quantitative agreement with Monte Carlo simulation results on regular lattices have been observed \cite{Buzano,tiago10}.

To solve the $k$NN mixtures on Husimi lattices, we proceed as usual by defining partial partition functions (ppf's) according to the state of the root site of an elementary cube. As shows Fig. \ref{fig1}f, in general, for each sublattice the root site can be empty ($j=\varnothing$), occupied by a 0NN ($j=0$), a 1NN ($j=1$) or a 2NN particle ($j=2$). In the 0NN-1NN model only $j=\varnothing$, $0$ and $1$ are allowed and since there are two sublattices one has a total of six possible states for the root site (and, so, six ppf's). The situation worsen in the 1NN-2NN mixture, where $j=\varnothing$, $1$ and $2$ are allowed and there are eight sublattices, totaling twenty-four states. 

Let us now consider the operation of attaching seven cubes to the vertices of another cube, with exception of the root site. This give us a 1-generation subtree. Repeating this process by attaching the root sites of seven subtrees to the vertices of another new cube, we can build a $(M+1)$-generation subtree from seven $M$-generation ones. So, if we keep the root site of the new cube in a given state $s$ and sublattice $g$, and sum over all the possible ways of attaching the seven $M$-generation subtrees to it, we obtain the ppf $g'_s$ in generation $(M+1)$ as a function of the ppf's $g_j$ in generation $M$, with $s,j=\varnothing,0,1$ and $g=a,b$ in the 0NN-1NN case and $s,j=\varnothing,1,2$ and $g=a_1,a_2,\ldots,d_1,d_2$ in the 1NN-2NN model. Therefore, we obtain six (twenty-four) recursion relations (RRs) for the ppf's of the 0NN-1NN (1NN-2NN) mixture. These RRs are presented in the appendix. Since they diverge in the thermodynamic limit ($M \rightarrow \infty$), we will work with ratios of them. From the RRs for the ppf's, we may define four (sixteen) RRs for the ratios in the 0NN-1NN (1NN-2NN) model, whose definitions are also presented in the appendix. The stable (real and positive) fixed points of these last RRs gives us the stable thermodynamic phases of the models. To determine the stability limits of a given phase, we calculate the Jacobian matrix at the related fixed point. Wherever the largest eigenvalue ($\Lambda$) of this matrix is smaller than 1, the fixed point is stable, as well as the corresponding thermodynamic phase. The condition $\Lambda = 1$ defines the stability limits.

The partition function ($Y$) of the models can be obtained, similarly to the ppf's, by attaching the root sites of eight subtrees to the eight vertices of a central cube. For the 0NN-1NN mixture, it can be written in a compact form, e.g, as 
\begin{equation}
Y = a_{\varnothing} a'_{\varnothing} + z_0 a_{0} a'_{0} + z_1 a_{1} a'_{1} = (a_\varnothing b_\varnothing)^4 y,
\label{eqY}
\end{equation}
where $y$ depends only on the activities and ratios $A_j=a_j/a_{\varnothing}$ and $B_j=b_i/b_{\varnothing}$, with $j=0,1$ (see the appendix) and, so, it attains a constant value in the thermodynamic limit. Using Eqs. \ref{RRs0NN1NN} in appendix to write down the expanded expression for $y$, the densities of small and large particles at the central cube, in sublattice $A$, are given respectively by 
\begin{equation}
\rho_0^{(A)} = \frac{A_0}{8Y} \frac{\partial Y}{\partial A_0} \quad \text{and} \quad \rho_1^{(A)} = \frac{A_1}{8Y} \frac{\partial Y}{\partial A_1}.
\label{eqDensities}
\end{equation}
By replacing $A$ to $B$ one obtains the densities in sublattice $B$. Thence, $\rho_j = \rho_j^{(A)} + \rho_j^{(B)}$ gives the total density of small ($j=0$) and large ($j=1$) particles at the central cube in the 0NN-1NN mixture. We can obtain $Y$ and the densities in a similar fashion for the 1NN-2NN system, as devised in the appendix.

The bulk free energy (per site) of each phase of the models can be calculated following the ansatz proposed by Gujrati \cite{Gujrati}, and discussed in detail for Husimi lattices built with cubes in Ref. \cite{Nathann19}. For the 0NN-1NN model it reads
\begin{equation}
 \phi_b = -\frac{1}{8} \ln \left[ \frac{\left( A_{\varnothing} B_{\varnothing} \right)^4}{y^{6}} \right],
\end{equation}
where $A_\varnothing=a'_\varnothing/a_\varnothing^3 b_\varnothing^4$ and $B_\varnothing=b'_\varnothing/a_\varnothing^4 b_\varnothing^3$. The equivalent expression for the 1NN-2NN mixture is presented in the appendix. In general, in regions of the parameter space where two or more phases are stable, the equality of their free energies will define the points,  lines, or surfaces of coexistence (where the first-order transitions take place). Since each lattice site occupies a volume $v_0=a^3$, the pressure (in our grand canonical formalism) is given by $P=-\phi/a^3$.

\section{Thermodynamic behavior of the 0NN-1NN mixture}
\label{secRes0NN1NN}

Before discussing the 0NN-1NN system, it is interesting to analyze the pure 1NN lattice gas. Once $z_0=0$ in this case, one has $A_0=A_1=A$ and $B_0=B_1=B$, so that only two recursion relations (RRs, $A$ and $B$) have to be analyzed (see the appendix). For small activity ($z_1 \leq 0.8298$) only a homogeneous, disordered fluid phase with $A=B$ is stable, while for large activity ($z_1 \geq 0.8298$) there are two equivalent fixed points, where $A=R>B=r$ or $A=r<B=R$. These last ones correspond to the two possible states of the ordered solid ($S1$) phase. Since the stability limits of both fluid and $S1$ phases coincide at $z_{1,c}=0.8298$ this is a critical point and there is a continuous fluid-solid transition in this model. This is consistent with several numerical studies of the 1NN model on the cubic lattice, where a critical point located at $z_1 \approx 1.05$ has been found \cite{HB,Yamagata,Panagiotopoulos}, which is a bit larger than the value from our mean-field approximation, as expected. In agreement with this, the critical density in the cubic lattice is given by $\rho_{1,c} \approx 0.21$ \cite{Panagiotopoulos,Gaunt}, while here one finds a slightly smaller value $\rho_{1,c} = 0.1762$. 

\begin{figure}[!t]
\includegraphics[width=8.cm]{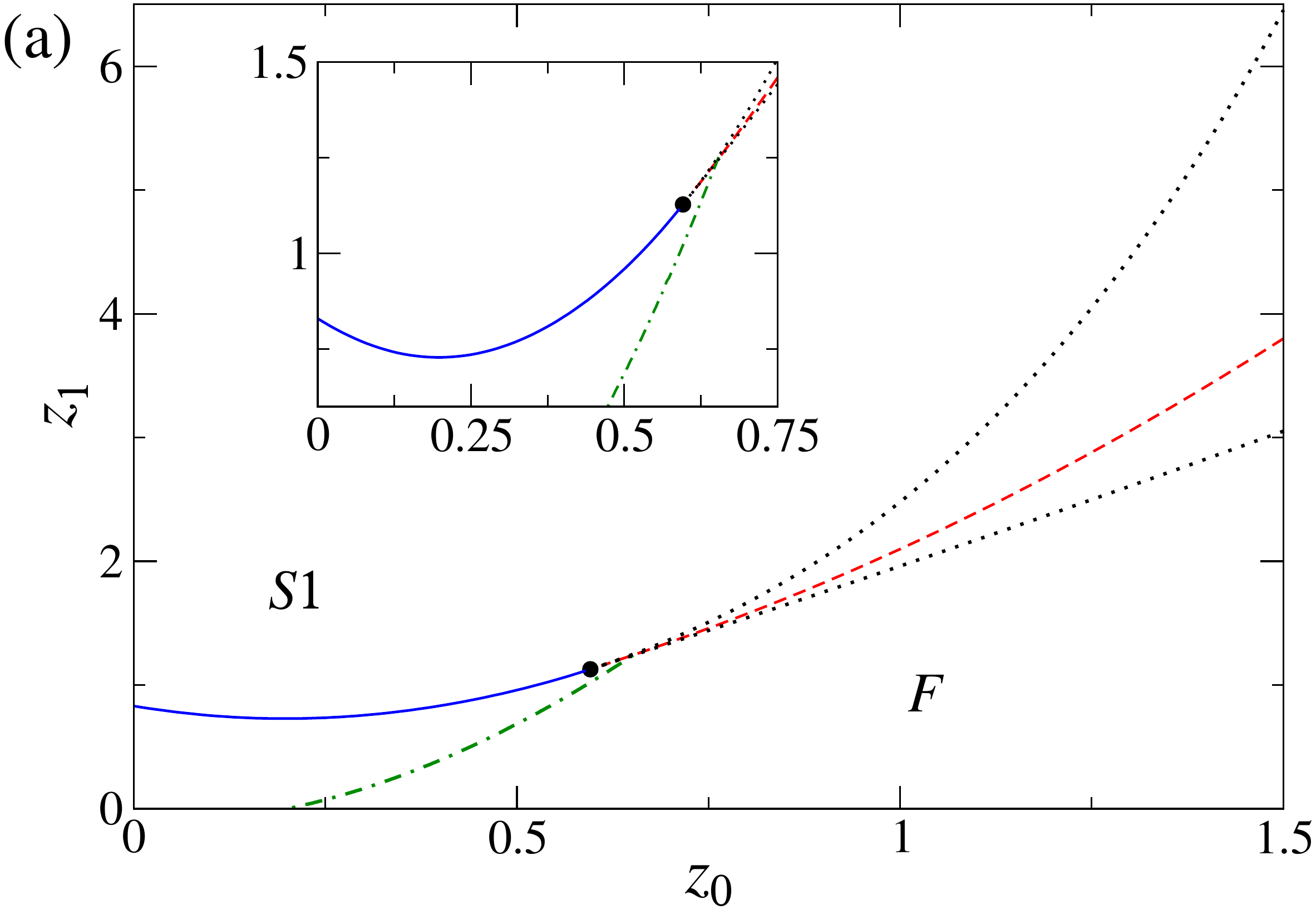}
\includegraphics[width=8.cm]{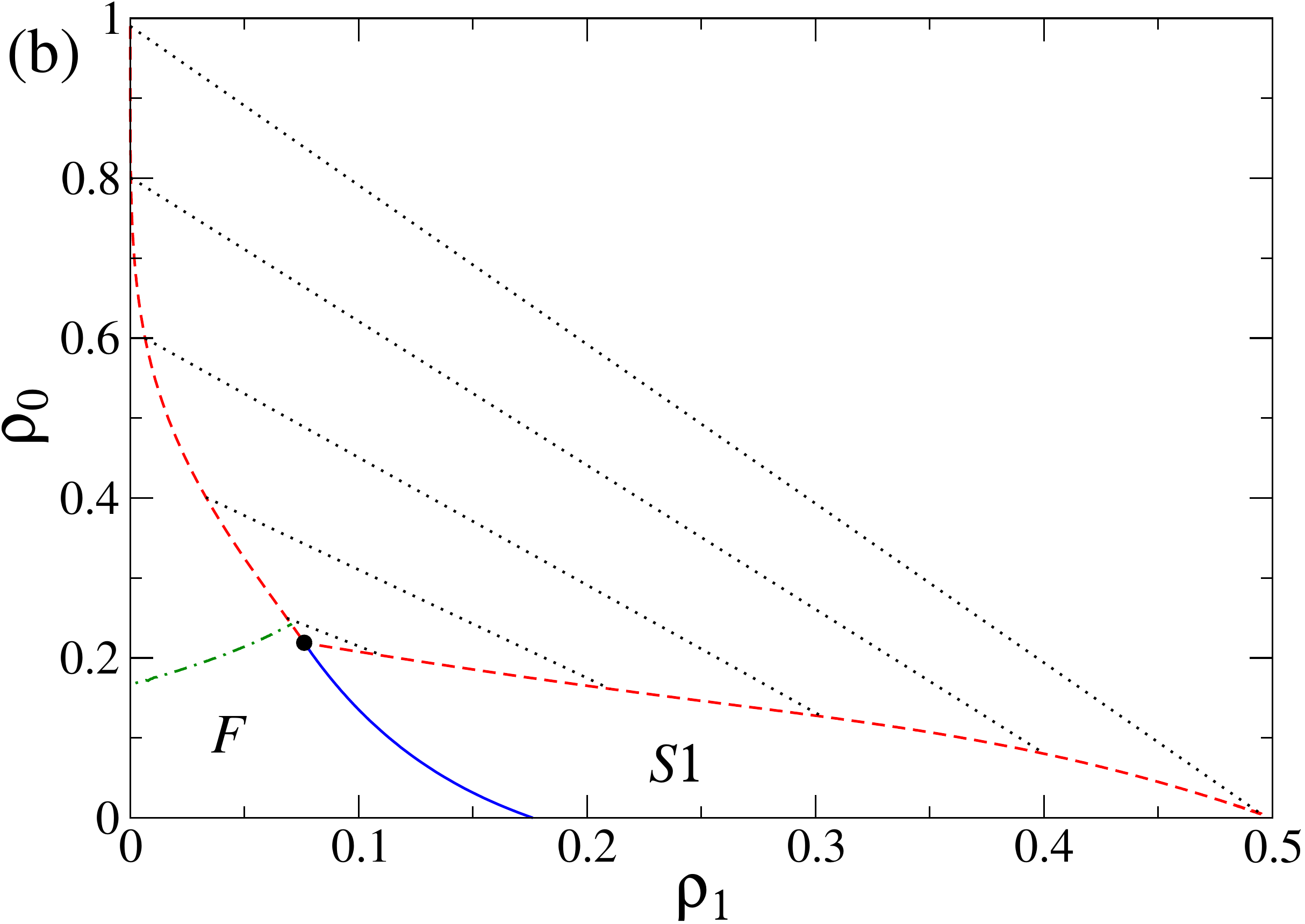}
\includegraphics[width=8.cm]{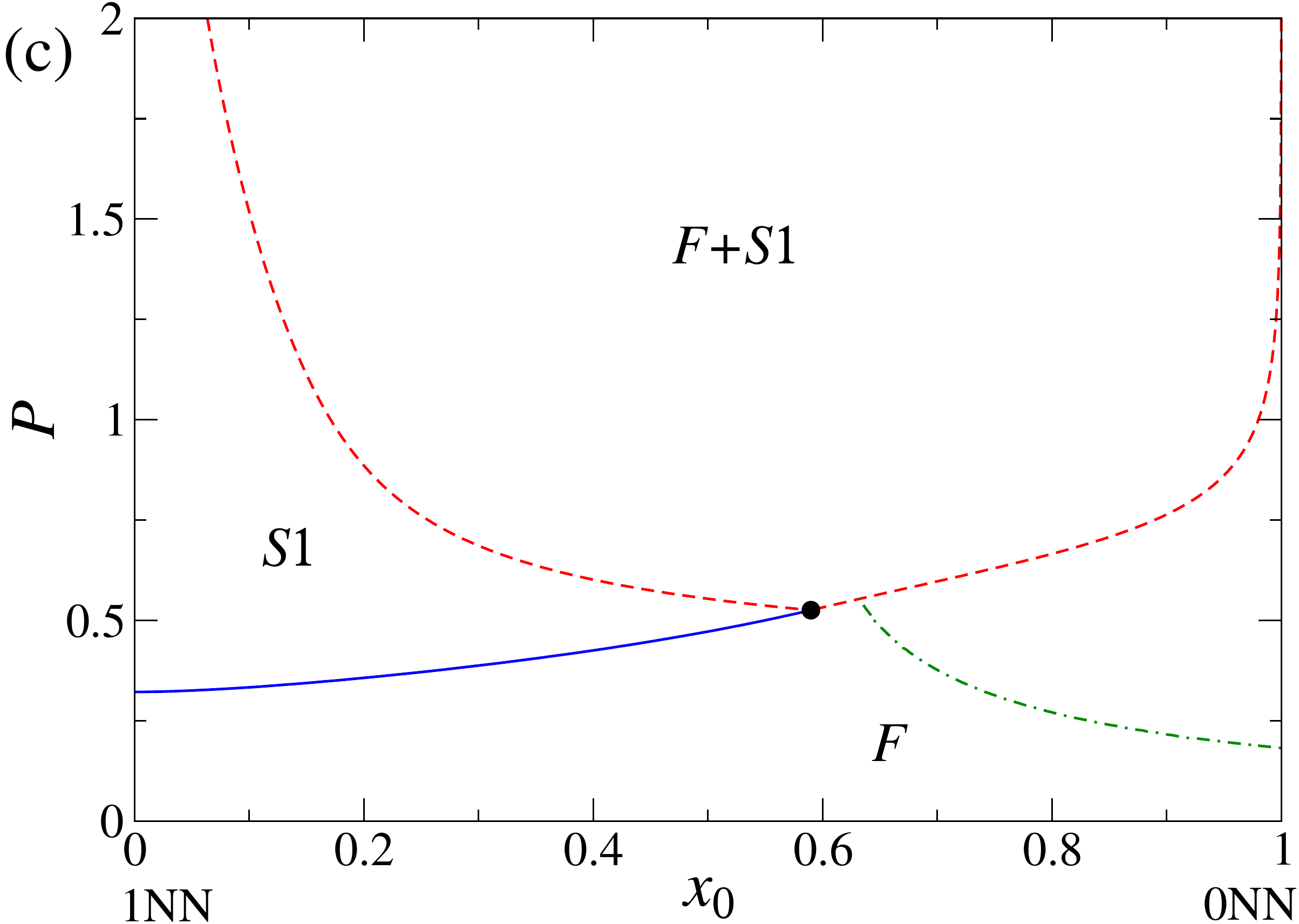}
\caption{Phase diagrams for the 0NN-1NN mixture in (a) activities ($z_1,z_0$), (b) densities ($\rho_0,\rho_1$) and (c) pressure-composition ($P,x_0$) spaces. In all panels, the solid (blue) and dashed (red) lines indicated the continuous and discontinuous transitions loci, respectively. The dash-dotted (green) lines are the LMDs. The tricritical point (TC) is indicated by the circle. The dotted lines in (a) are the spinodals, while in (b) they are tie lines. The inset in (a) highlights the region around the minimum in the critical line and the TC point.}
\label{fig2}
\end{figure}

By including the 0NN particles in the system, we observe that $A_0\neq A_1$ and $B_0\neq B_1$, so that now we have to deal with four RRs for the ratios. In the fluid phase $A_j=B_j$, while in the solid phase $A_j > B_j$ when sublattice $A$ is the one more populated, for $j=0,1$. For small activity (and so density) of point particles, one still find a continuous fluid-solid transition, which becomes discontinuous for large $z_0$. Namely, for small $z_0$ the stability limits of the fluid and solid phases are coincident, giving rise to a critical line, but for  $z_0 > z_{0,TC}=0.5958$ (and $z_1 > z_{1,TC}=1.1277$) they become different, yielding a coexistence region (see Fig. \ref{fig2}a). From the equality of the free energies of both phases, one finds a first-order transition line which starts at  $(z_{0,TC},z_{1,TC})$ and extends to $z_0,z_1 \rightarrow \infty$. In such limit, the fixed point of the $S1$ phase is given, e.g, by $A_0=A_1=1$ and $B_0=B_1=0$, whilst the $F$ phase is characterized by $A_0=B_0= 1$ and $A_1=B_1=0$. By calculating the free energy with these limiting values, we find that the $F$-$S1$ coexistence line is given by
\begin{equation}
 z_1 \simeq z_0^2 + z_0,
\end{equation}
for large $z_0$ and $z_1$. Therefore, for $z_0 \rightarrow \infty$ we have $z_1 \approx z_0^2$, which is consistent with the fact that effectively one 1NN particle occupies the volume of two 0NN ones.

The critical line starts at $(z_{0},z_{1})=(0, 0.8298)$ and tangentially meet the coexistence line at $(z_{0,TC},z_{1,TC})$, demonstrating that this is a tricritical point. In contrast with the coexistence line, which is a monotonic increasing function of $z_0$ [in ($z_0,z_1$) space], the critical line initially decreases with $z_0$, having a slope $d z_1/d z_0|_{z_0 \rightarrow 0} = -1$. Then, it passes through a minimum at $(z_{0},z_{1})=(0.1974, 0.7284)$ and finally increases towards the tricritical point. See Fig. \ref{fig2}a. Such initial decreasing of the critical line shows that a small density of small particles facilitates the ordering of the large ones, which is certainly due to an effective attractive depletion interaction among the 1NN particles. We remark that a very similar behavior have been found for small $z_0$'s in the solution of the 0NN-2NN mixture on a Husimi lattice built with cubes \cite{Nathann19}, although in such case the fluid-solid transition is discontinuous. Moreover, in studies of the 0NN-1NN model on the square \cite{Jim01,tiago15} and Bethe lattice \cite{tiago11} a minimum in the critical fluid-solid transition line has also been found. Actually, the full phase diagram for the 0NN-1NN mixture on these lattices is qualitatively identical to the one found here, with a continuous and a discontinuous transition line meeting at a tricritical point \cite{Jim01,tiago15,tiago11}.

This is true also for the diagram in density space, as displayed in Fig. \ref{fig2}b, where the critical line starts at $(\rho_{0},\rho_{1})=(0,0.1762)$ and ends at the tricritical point, which is located at $(\rho_{0,TC},\rho_{1,TC})=(0.2190,0.0761)$. There, one can see that in general the solid phase is featured by a small density of 0NN particles ($\rho_0 \leq \rho_{0,TC}$), whereas the opposite happens in the fluid phase (where $\rho_1 \leq 0.1762$). For large $z_0$ (and $z_1$), using the asymptotic behaviors discussed above for the RRs and coexistence line, it is straightforward to demonstrate that $\rho_0^{(S1)} \approx \frac{1}{2z_0}$ and $\rho_1^{(S1)} \approx \frac{1}{2} - \frac{1}{2z_0}$ in the solid phase, while $\rho_0^{(F)} \approx 1 - \frac{1}{z_0}$ and $\rho_1^{(F)} \approx \frac{1}{z_0^5}$ in the fluid one. Thereby, as $z_0 \rightarrow \infty$ one obtains the expected densities in the full occupancy limit: $\rho_0^{(S1)} \rightarrow 0$ and $\rho_1^{(S1)} \rightarrow 1/2$, and $\rho_0^{(F)} \rightarrow 1$ and $\rho_1^{(F)} \rightarrow 0$, as indeed observed in Fig. \ref{fig2}b.

The fluid-solid demixing transition is also clearly observed in the pressure ($P$) versus composition ($x_0$) phase diagram displayed in Fig. \ref{fig2}c. Here, the molar fraction of 0NN particles was defined as $x_0=\rho_0/\rho_T$, where $\rho_T = \rho_0 + 2\rho_1$ is the total density of particles. In this phase diagram one sees that for small pressures ($P < 0.3219$) only the fluid phase is stable, regardless the composition $x_0$. For larger $P$, up to the tricritical point, which is located at $(x_{0,TC},P_{TC})=(0.5898,0.5258)$, there is a continuous $F$-$S1$ transition line. For $P>P_{TC}$ a coexistence region is observed. We notice that in the limit of large $P$ (corresponding to large $z_0$ and $z_1$) the molar fraction in the fluid and solid phase tend to $x_0=1$ and $x_0=0$, respectively. In this limit, it is simple to demonstrate that $P$ diverge as $P^{(F)} \sim -\ln (1-x_0)$ and $P^{(S1)} \sim -\ln(x_0)$.

Finally, let us notice that, similarly to what happens in the 0NN-1NN mixture on the square lattice, a density anomaly is also observed in our cubic approximation, which is characterized by minima in the isobaric curves of the total density of particles $\rho_T$ against $z_0$ (or $z_1$). A behavior analogous to the ones displayed, e.g, in Figs. 5a and 6 from Refs. \cite{Nathann19} and \cite{tiago15}, respectively. The ($z_0,z_1$) coordinates of these minima gives rise to a line of minimum density (LMD), which is also shown in the phase diagram of Fig. \ref{fig2}a. Such line starts inside the fluid phase at $z_0 \approx 0.2$ as $z_1 \rightarrow 0$, which is a result consistent with the one analytically predicted in the solution of the model on the Bethe lattice \cite{tiago11}, where the LMD was found to start at $z_0=1/(q-1)$, with $q$ being the lattice coordination ($q=6$, here). Then, the LMD crosses the stable and metastable fluid regions, ending at the spinodal of this phase. The LMD in other thermodynamic variables are also depicted in the Figs. \ref{fig2}b and \ref{fig2}c, where the same behavior is observed, as expected. We remark that such behavior is somewhat different from that found on the square lattice, where the LMD seems to end at the tricritical point \cite{tiago11,tiago15}. On the other hand, it is similar to the LMD found for the 0NN-2NN mixture on the Husimi lattice built with cubes, which also ends inside the region where the fluid is metastable \cite{Nathann19}, indicating that this can be a general feature of the LMDs of such mixtures in the cubic lattice. Regardless the location of their end points, in all these systems the LMDs divide the fluid phase in two regions: a regular one, for small $z_0$, $\rho_0$ and $x_0$, where $\partial \rho_T/\partial z_0 <0$, as expected; and the anomalous region, for large $z_0$, $\rho_0$ and $x_0$, where $\partial \rho_T/\partial z_0 >0$.


\section{Thermodynamic behavior of the 1NN-2NN mixture}
\label{secRes1NN2NN}

\begin{figure}[!b]
\includegraphics[width=8.cm]{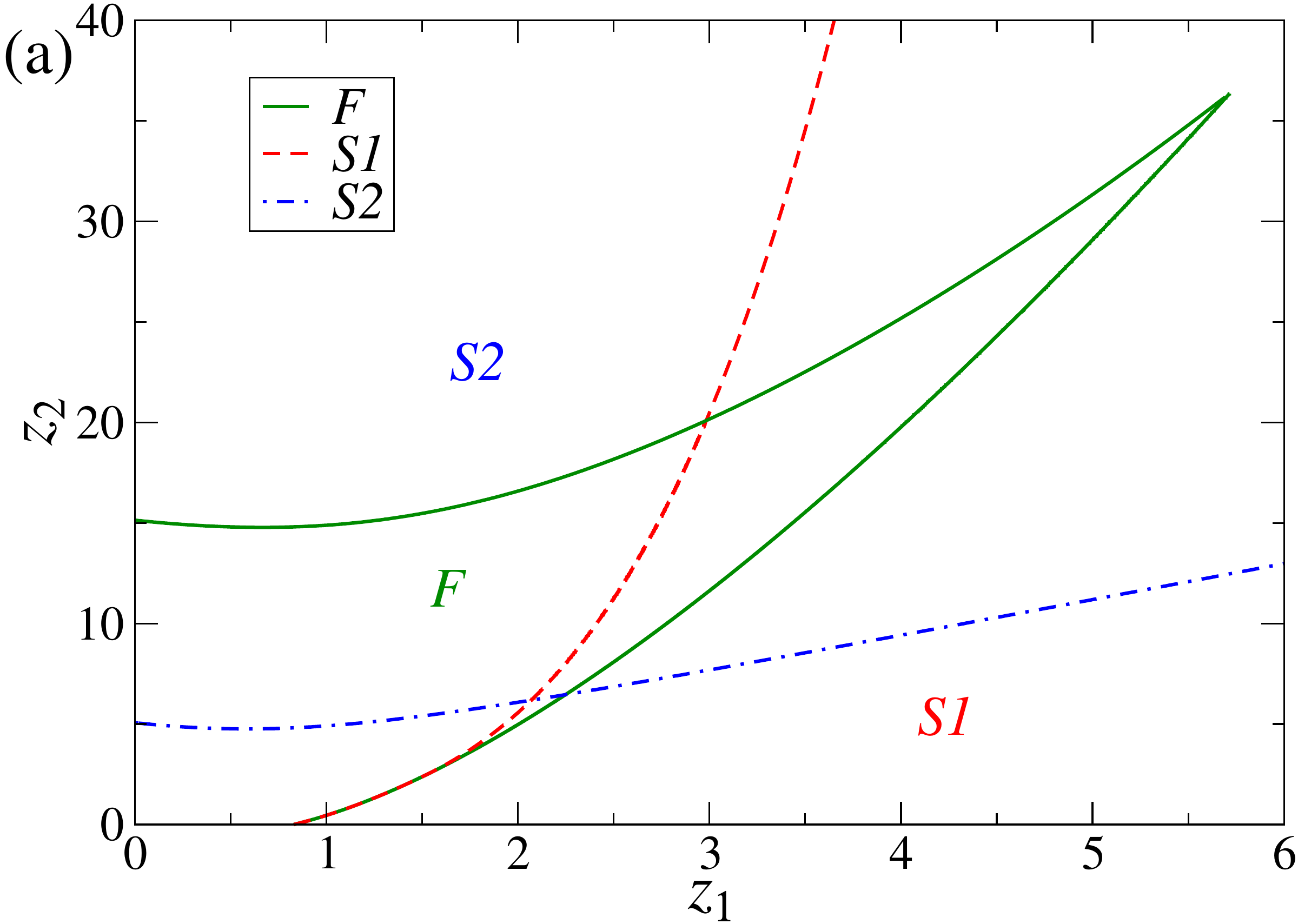}
\includegraphics[width=8.cm]{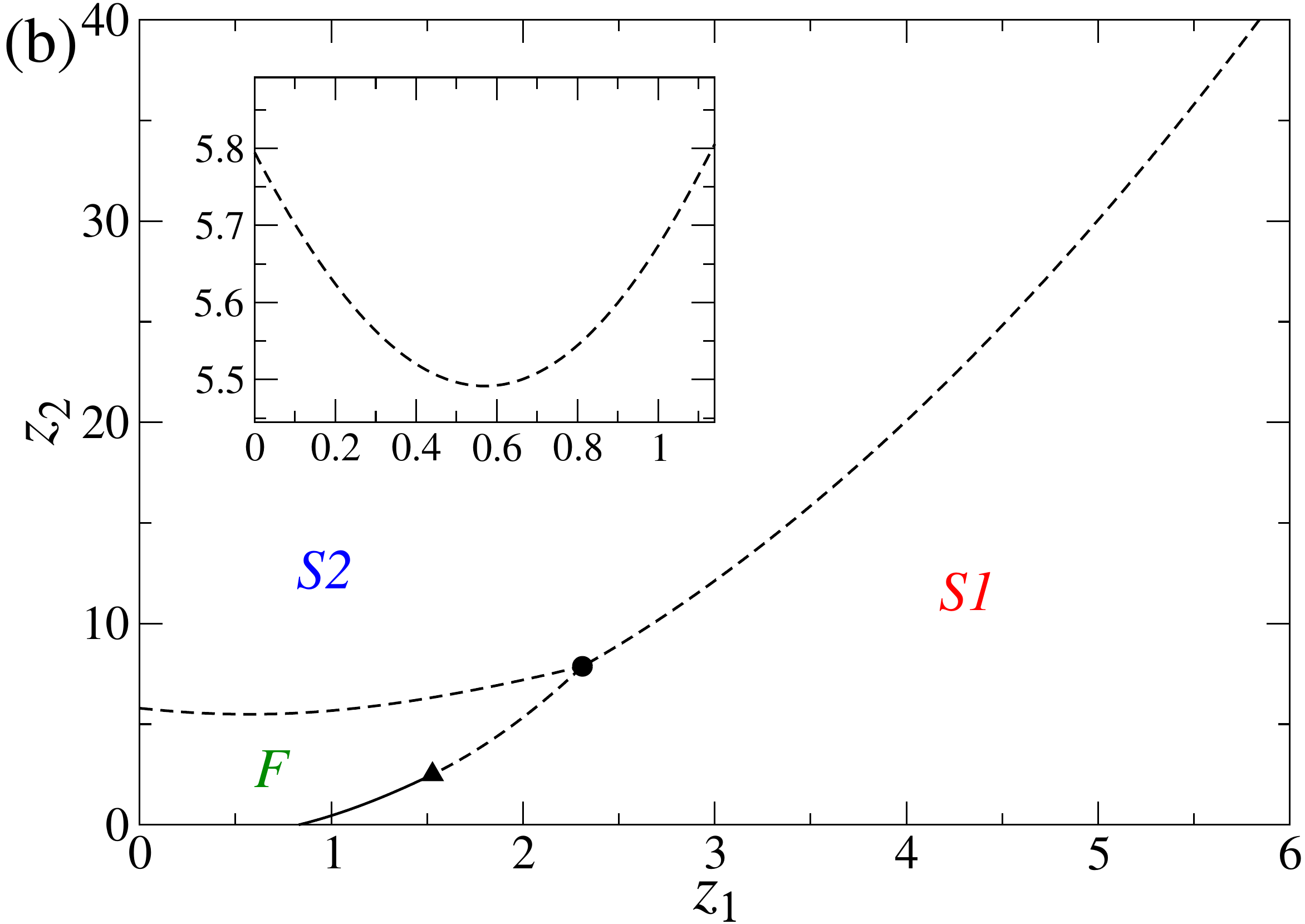}
\caption{(a) Stability limits of the three stable (fluid, $S1$ and $S2$) phases found in the 1NN-2NN mixture. (b) Phase diagram of this model in variables $z_1,z_2$. In (b), the solid and dashed lines represent continuous and discontinuous transitions, respectively. The triple point is indicated by a circle and the tricritical point by a triangle. The inset highlights the region around the minimum in the $F$-$S2$ coexistence line.}
\label{fig3}
\end{figure}

Now, we investigate the 1NN-2NN model. Let us starting noticing that three stable phases are found in this system: \textit{i)} the isotropic, disordered fluid ($F$) phase, for which the RRs (at the fixed point) have the form $A_{i,j}=B_{i,j}=C_{i,j}=D_{i,j}$ for $i=1,2$ and $j=1,2$; \textit{ii)} the solid ($S1$) phase, associated with the ordering of 1NN particles, whose fixed point is featured by, e.g, $A_{1,j}=B_{1,j}=C_{1,j}=D_{1,j} > A_{2,j}=B_{2,j}=C_{2,j}=D_{2,j}$, for $j=1,2$, when the sublattices indexed by $1$ (see Fig. \ref{fig1}e) are the more populated ones; and \textit{iii)} the solid ($S2$) phase, associated with the ordering of 2NN particles, where we find, e.g, $A_{i,j} > B_{i,j}=C_{i,j}=D_{i,j}$, for $i=1,2$ and $j=1,2$, when the sublattices labeled by $A$ are the ones more occupied. Although the notations are different, the fixed points' symmetries for the fluid and $S1$ phases are the same from the previous section. Since for the pure 1NN model ($z_2=0$) a continuous fluid-solid ($F$-$S1$) transition is found (as discussed above), we might expect to find a continuous $F$-$S1$ transition line for small $z_2$. On the other hand, for the pure 2NN model ($z_1=0$) a discontinuous fluid-solid ($F$-$S2$) transition takes place, as observed by Panagiotopoulos \cite{Panagiotopoulos} in Monte Carlo simulations on the cubic lattice and by us in the cubic Husimi lattice approximation \cite{Nathann19}. Therefore, for small $z_1$ a first order $F$-$S2$ transition line is expected. Such behaviors are indeed found here for the 1NN-2NN mixture, as confirmed in Fig. \ref{fig3}. There, the stability limits of the three stable phases are shown in panel (a), where one sees that for small $z_2$ the spinodals of the fluid and $S1$ phases indeed coincide, up to $(z_{1,TC},z_{2,TC})=(1.5273,2.5016)$, after which they become different, giving rise to a coexistence region between these phases. The spinodal of the $S2$ phase, on the other hand, never coincides with the ones for the fluid and $S1$ phases, with exception of few points where they cross each other, indicating that such phases are always separated by first order transition lines.

\begin{figure}[!t]
\includegraphics[width=8.cm]{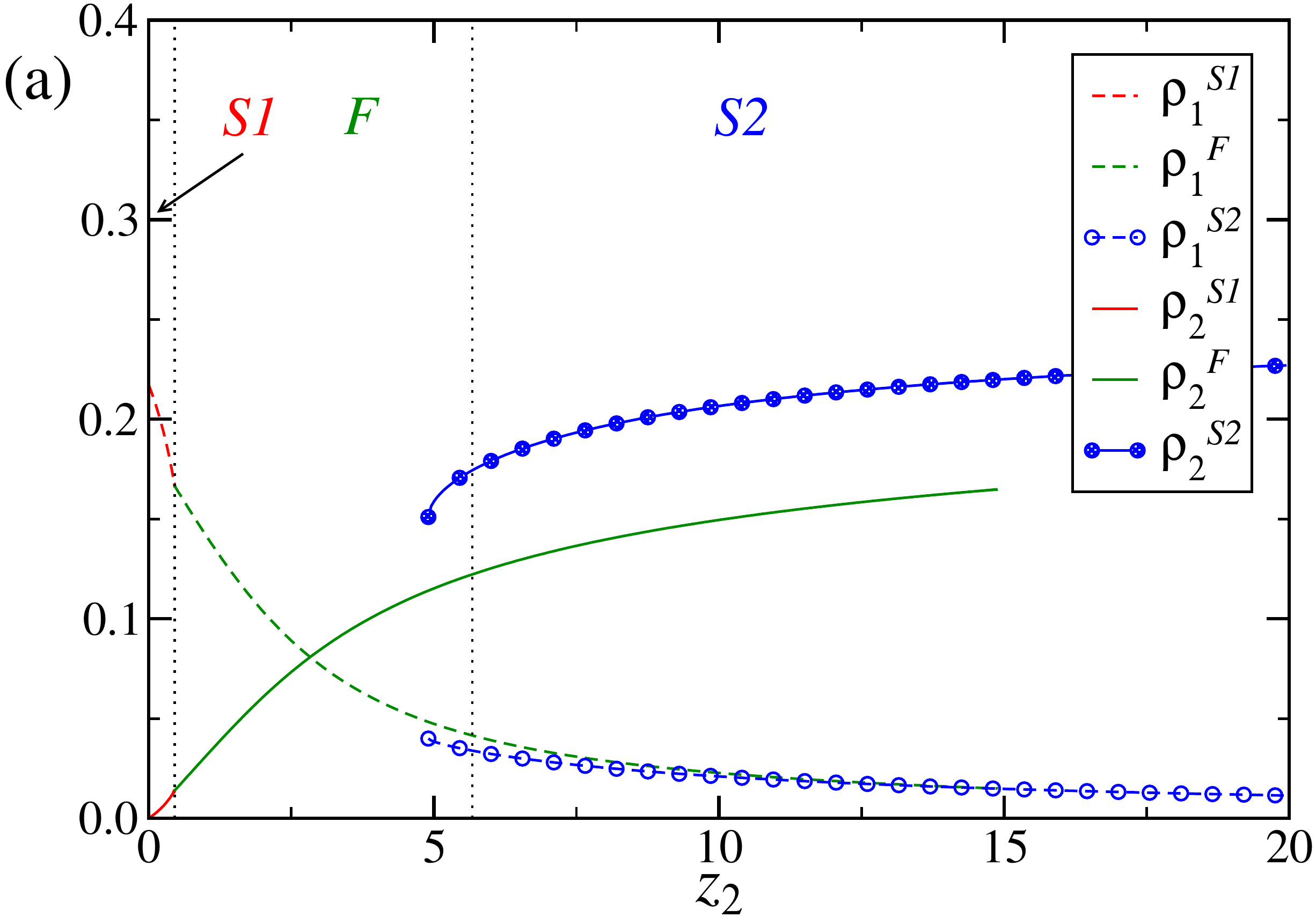}
\includegraphics[width=8.cm]{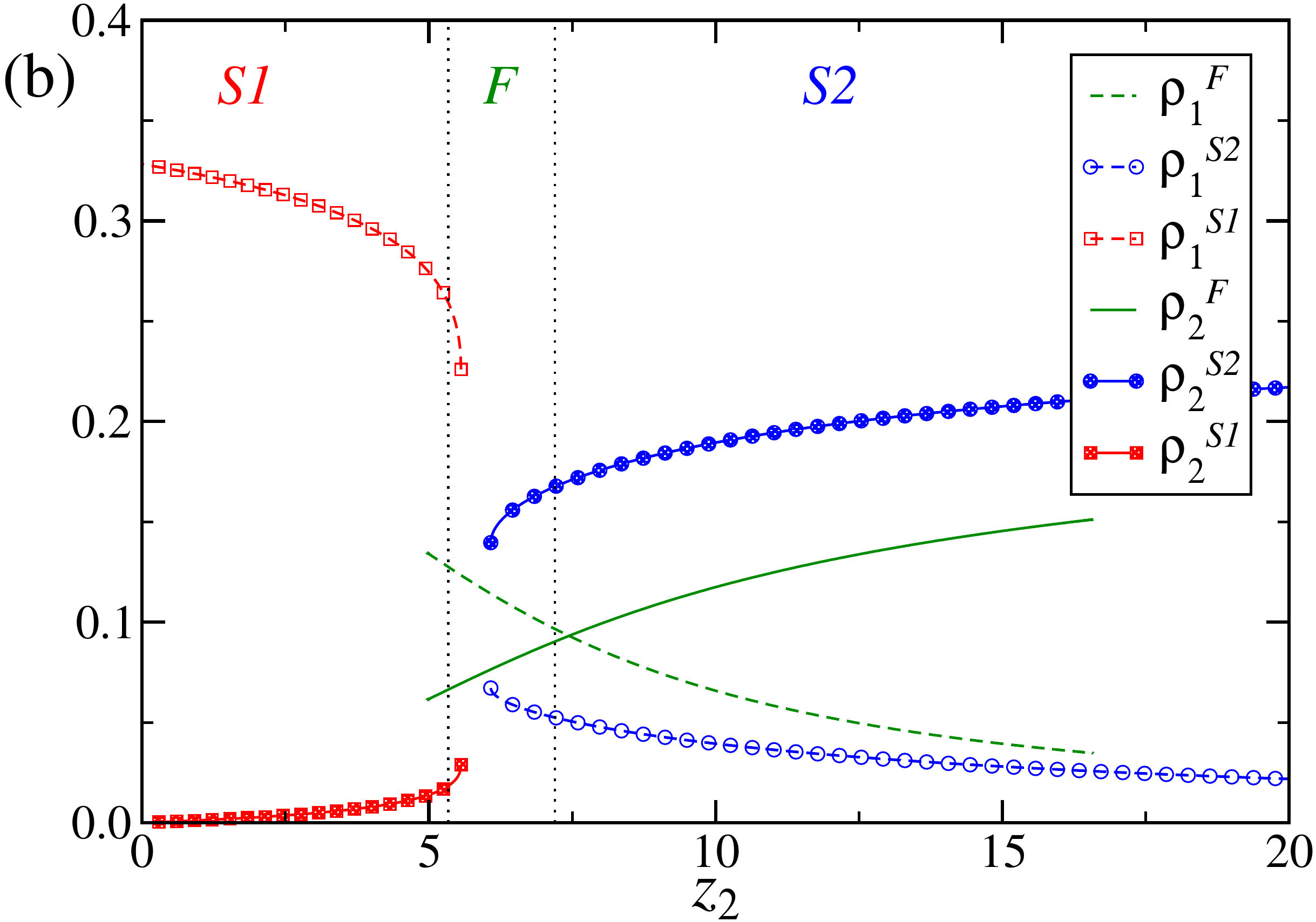}
\includegraphics[width=8.cm]{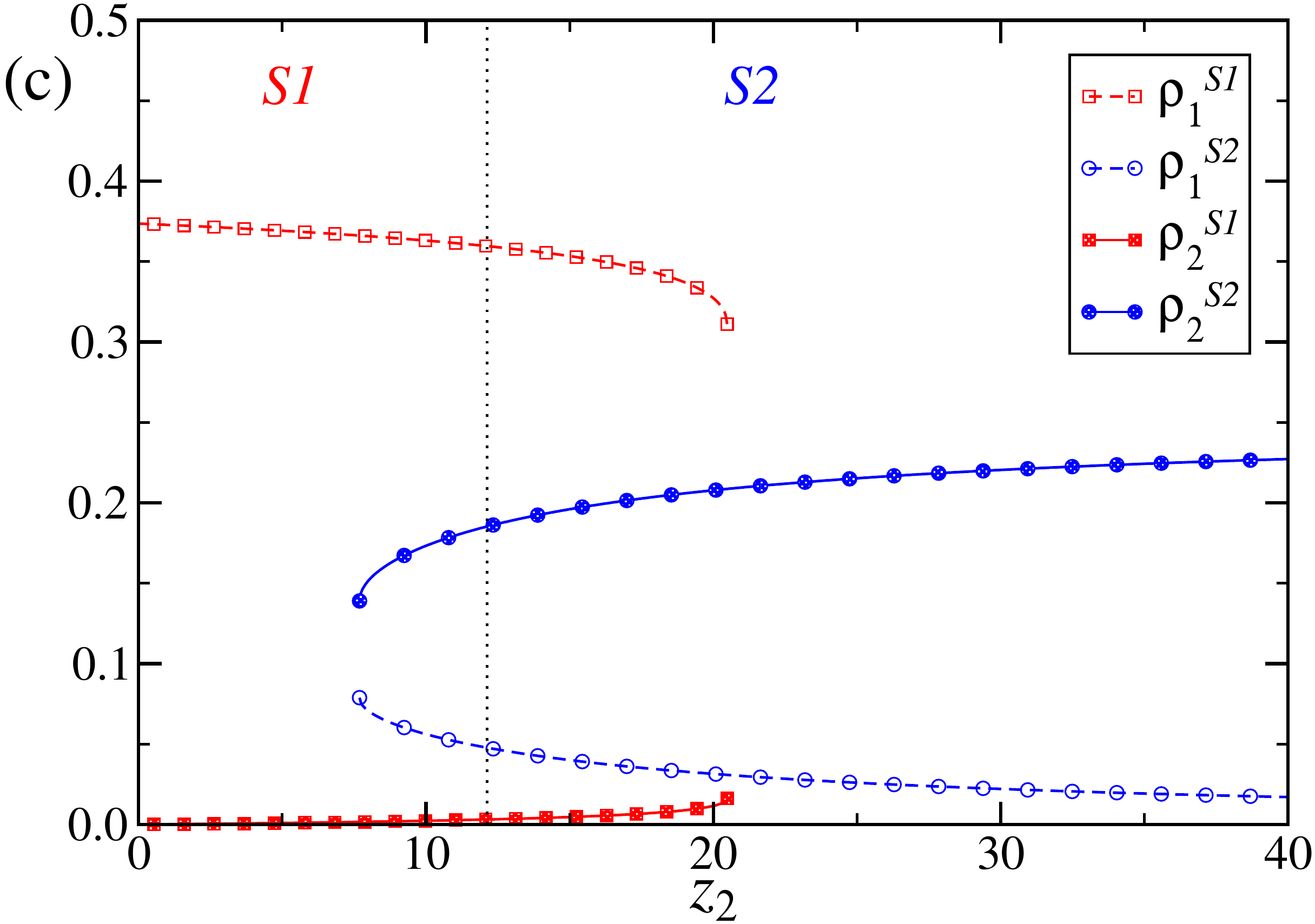}
\caption{(a) Densities of small $\rho_1$ and large $\rho_2$ particles in the indicated phases, for the 1NN-2NN mixture, as functions of $z_2$ for (a) $z_1=1$, (b) $z_1=2$ and (c) $z_1=3$. The vertical lines are the transition loci between the indicated phases.}
\label{fig4}
\end{figure}

In fact, from the equality of the bulk free energies, three coexistence ($F$-$S1$, $F$-$S2$ and $S1$-$S2$) lines are obtained, all of them meeting at a triple point, located at $(z_{1,TP},z_{2,TP})=(2.3102,7.8746)$. See Fig. \ref{fig3}b. The $F$-$S1$ coexistence line tangentially meet the $F$-$S1$ critical line at ($z_{1,TC},z_{2,TC}$), showing that this is a tricritical point. The $F$-$S2$ coexistence line starts at $z_1=0$ (and $z_2=5.7932$) and ends at the triple point. In between, however, it presents a minimum located at $(z_1,z_2)=(0.5702,5.4907)$, as shows the insertion in Fig. \ref{fig3}b. Therefore, initially this line decreases, indicating that a small fraction of 1NN particles facilitates the ordering of the larger 2NN ones. This is similar to the effect of the 0NN particles on the ordering of 1NN ones discussed in the previous section or in Refs. \cite{tiago11,tiago15} for the square lattice, as well as on the ordering of 2NN particles, as demonstrated in \cite{Nathann19} for the 0NN-2NN mixture. Interestingly, in all these systems the initial slope of the decreasing critical or coexistence lines is $d z_k/d z_0|_{z_0\rightarrow 0}=-1$ (with $k=1,2$), and here one also finds $d z_2/d z_1|_{z_1 \rightarrow 0}=-1$.

The $S1$-$S2$ coexistence line starts at the triple point and extends to $z_1,z_2\rightarrow\infty$. In such limit, one finds that all RRs vanish in $S1$ phase, with exception, e.g, of $A_{1,1}=B_{1,1}=C_{1,1}=D_{1,1}= 1$, while in the $S2$ phase one has, e.g, $A_{i,j} = 1$ and $B_{i,j}=C_{i,j}=D_{i,j} = 0$, for $i=1,2$ and $j=1,2$. Inserting these limiting values in the free energies, we find the  $S1$-$S2$ coexistence line as
\begin{equation}
 z_2 \simeq z_1^2 + z_1,
\end{equation}
for large $z_1$. So, for $z_1 \rightarrow \infty$ one obtains $z_2 \approx z_1^2$, which is again consistent with the fact that effectively two 1NN particles occupy the same volume of a 2NN one.

\begin{figure}[!t]
\includegraphics[width=8.cm]{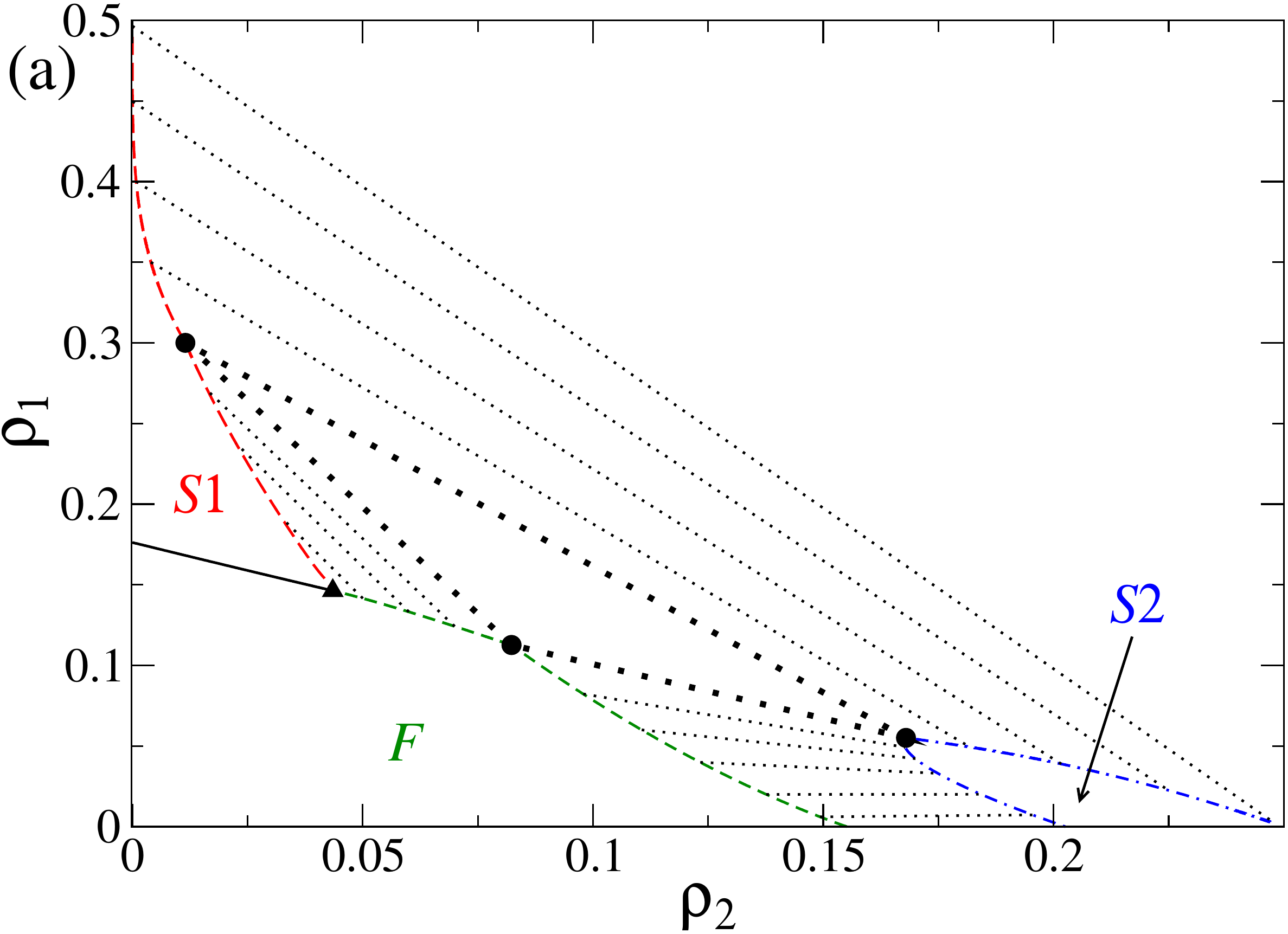}
\includegraphics[width=8.cm]{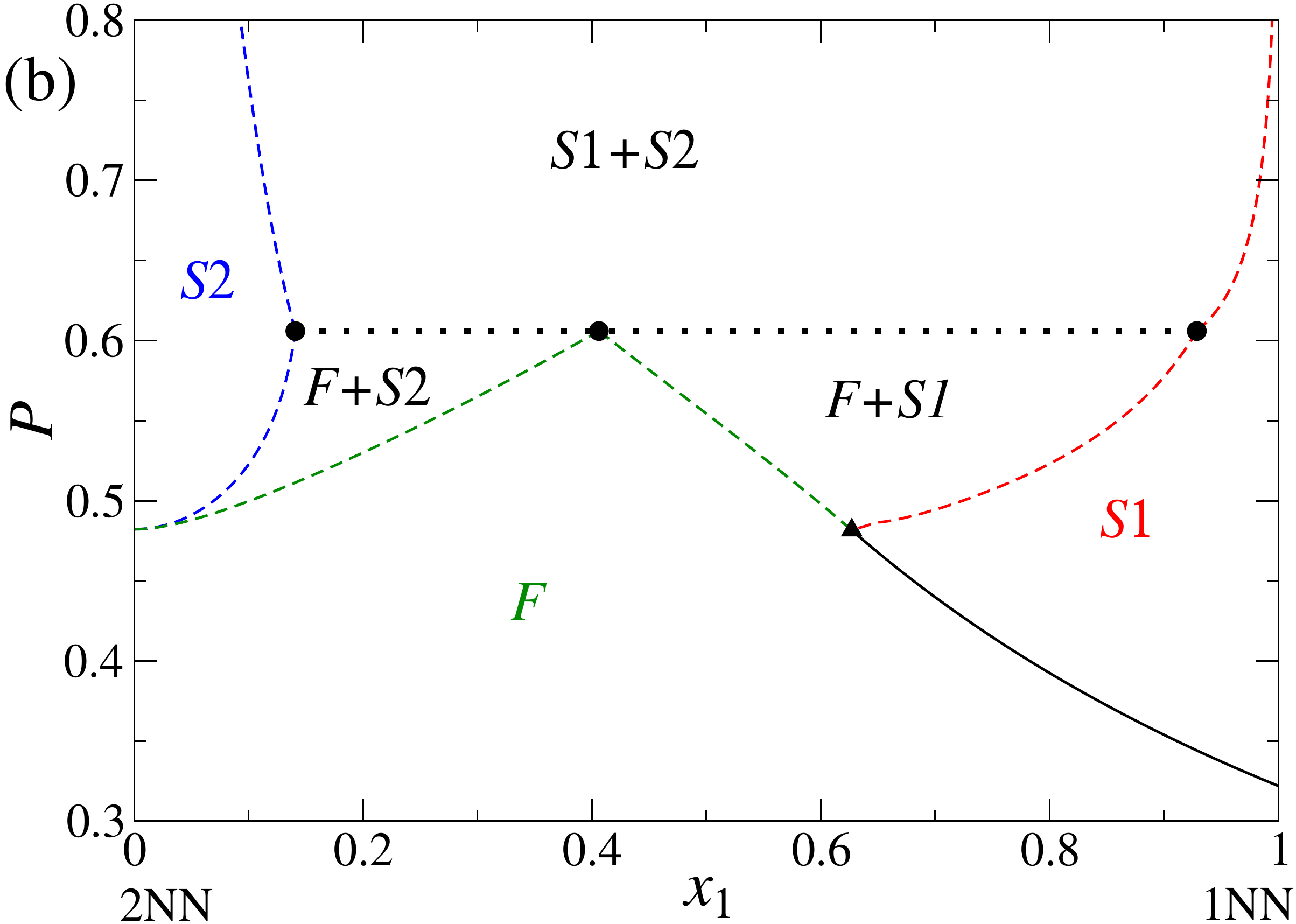}
\caption{Phase diagrams for the 1NN-2NN mixture in (a) density ($\rho_1,\rho_2$) and (b) pressure-composition ($P,x_1$) space. In both panels the solid and dashed lines denotes the continuous and discontinuous transitions, respectively. The tricritical point is indicated by the triangle, while the circles represent the triple point, which are connect by thicker dotted lines. The thin dotted lines in (a) are tie lines.}
\label{fig5}
\end{figure}

To let clear the differences among the three phases, specially between the two solid ones, it is worthy analyzing in detail the particle densities. Figures \ref{fig4}a-c display the densities $\rho_1$ and $\rho_2$ as functions of $z_2$ for three values of $z_1$, chosen such that all transition lines are crossed, where the continuous and discontinuous nature of the transitions are confirmed. As expected, for all phases and parameters $\rho_1$ decreases, while $\rho_2$ increases, with $z_2$. The $S1$ ($S2$) phase is always featured by a large density of 1NN (2NN) particles and a small density of 2NN (1NN) ones. In the fluid phase, on the other hand, the densities are more sensitive to the activities. For instance, at the $F$-$S2$ coexistence we find $\rho_1^{(F)}<\rho_2^{(F)}$ for $z_1=1$, but $\rho_1^{(F)}>\rho_2^{(F)}$ for $z_1=2$ (see Figs. \ref{fig4}a and \ref{fig4}b). These density behaviors are also confirmed in the phase diagram in the ($\rho_1,\rho_2$) space, which is depicted in Fig. \ref{fig5}a. In such diagram, the tricritical point is located at $(\rho_{1,TC},\rho_{2,TC})=(0.1463,0.0435)$, while at the triple point one has $\left( \rho_{1}^{(F)},\rho_{2}^{(F)} \right) = (0.1126,0.0823)$, $\left( \rho_{1}^{(S1)},\rho_{2}^{(S1)}\right)=(0.3000,0.0115)$ and $\left( \rho_{1}^{(S2)},\rho_{2}^{(S2)}\right)=(0.0550,0.1679)$. The $F$-$S1$ critical line starts at $(\rho_{1},\rho_{2})=(0.1762,0)$ and ends at the tricritical point. From the asymptotic behaviors discussed above for the RRs and the $S1$-$S2$ coexistence line in the limit of large $z_1$, it is easy to show that $\rho_1^{(S1)} \approx \frac{1}{2}-\frac{1}{2z_1}$ and $\rho_2^{(S1)} \approx \frac{1}{2z_1^4}$, while $\rho_1^{(S2)} \approx \frac{1}{4z_1}$ and $\rho_2^{(S2)} \approx \frac{1}{4}-\frac{1}{4 z_1}$. Hence, for $z_1 \rightarrow \infty$ we obtain $\rho_1^{(S1)} \rightarrow 1/2$ and $\rho_2^{(S1)} \rightarrow 0$, and $\rho_1^{(S2)} \rightarrow 0$ and $\rho_2^{(S2)} \rightarrow 1/4$, as expected and confirmed in Fig. \ref{fig5}a.

Figure \ref{fig5}b shows the phase diagram in pressure-composition ($P,x_1$) plane, where the molar fraction of 1NN particles was defined as $x_1=2\rho_1/\rho_T$, with $\rho_T = 2\rho_1 + 4\rho_2$ being the total density of particles in this model. For small pressures, specifically for $P < 0.3219$, only the fluid phase is stable, but above this point the $S1$ phase becomes also stable and both phases are separated by the critical line. This line ends at the tricritical point, located at $(P_{TC},x_{1,TC})=(0.4816,0.6270)$, above which a $F$-$S1$ coexistence appears. For higher pressures, the $F$-$S2$ and $S1$-$S2$ coexistences also show up. For instance, the $F$-$S2$ one starts at $(P,x_1)=(0.4822,0)$. The triple point, which separates the three regions where each kind of coexistence is the stable one, is located at $P_{TP}=0.6059$, where $x_1^{(S2)}=0.1408$,  $x_1^{(F)}=0.4061$ and  $x_1^{(S1)}=0.9286$. For large pressures the $S1$ line tends to $x_1 = 1$, while the $S2$ one goes to $x_1=0$, as expected. From the asymptotic behaviors discussed above for the RRs, coexistence line and densities, it is simple to demonstrate that $P$ diverges as $P^{(S1)} \sim -\ln(1-x_1)$ and $P^{(S2)} \sim -\ln(x_1)$. These behaviors are the same found in the previous section for the 0NN-1NN mixture and for the 0NN-2NN model in \cite{Nathann19}, indicating that such logarithmic divergences are universal in these systems.

Interestingly, in contrast with the other mixtures of $k$NN particles studied so far, the isobaric curves of the total density of particles ($\rho_T$) are always monotonic decreasing functions of $z_1$ or $z_2$ . Namely, no density anomaly exists in the 1NN-2NN mixture. This is certainly related to the fact that now the fluid phase is limited to a region of small activities and, so, of small densities. This contrasts with the 0NN-1NN mixture (discussed above and in previous studies related to square lattice \cite{tiago15,Jim01,tiago11}), as well as with the 0NN-2NN system \cite{Nathann19}, where the fluid phase extends to $z_0 \rightarrow \infty$ and, then, $\rho_0^{(F)},\rho_T^{(F)} \rightarrow 1$. Once the anomaly appears for relatively large densities of the small particles, this strongly suggests that it is absent in the 1NN-2NN case because it is preempted by the $F$-$S1$ transitions.

\section{Conclusions}
\label{secConc}

We have investigated two binary mixtures of (hard) $k$NN particles, which exclude up to their $k$th nearest neighbors and are discretized approximations for hard-spheres on the cubic lattice. More specifically, we have obtained the grand-canonical solution, on a Husimi lattice built with cubes, of the model with point particles (0NN) and 1NN ones; and for the mixture 1NN-2NN. The 0NN-1NN system displays the same thermodynamic behavior already observed for it on the square \cite{tiago15,Jim01,poland} and on the Bethe lattice \cite{tiago11}, with a disordered fluid ($F$) and a solid ($S1$) phase where 1NN particles tend to preferentially occupy one of two sublattices. Such phases are separated by a critical line and a first-order transition line, both meeting at a tricritical point. As demonstrated in Ref. \cite{tiago15}, the critical line and the tricritical point found for this mixture on the square lattice belong to the critical and tricritical Ising universality classes in 2D, respectively. Moreover, 3D Ising exponents were obtained by Heringa and Bl\"ote \cite{HB} for the pure 1NN model on the cubic lattice. These results suggest that the entire critical line, as well as the tricritical point might be also in the 3D Ising universality classes. Furthermore, in the Bethe lattice with coordination $q=6$, which is also a mean-field approximation for the cubic lattice, the tricritical point is located at $(z_{0,TC},z_{1,TC})=(0.5333,0.9957)$ \cite{tiago11}, which is smaller than the values found here: $(z_{0,TC},z_{1,TC})=(0.5958,1.1277)$. Since the Husimi lattice solution is an improved approach when compared with the Bethe lattice one, this strongly suggests that $(z_{0,TC},z_{1,TC})$ in the cubic lattice shall be a bit larger than the values found here. Numerical investigations of this mixture on the cubic lattice are necessary to confirm these things.

For the 1NN-2NN mixture, a disordered fluid ($F$) and the ordered $S1$ phases are still present in the phase diagram and, once again, they are separated by a critical and a coexistence line, which meet at a tricritical point. Nonetheless, there exists also a second solid ($S2$) phase, which is featured by the ordering of 2NN particles into one of four sublattices. Such phase is separated from the $F$ and $S1$ phases by first-order transition lines. Therefore, beyond the fluid-solid ($F$-$S1$) demixing also observed in the 0NN-1NN mixture, in the 1NN-2NN case there exist also another fluid-solid ($F$-$S2$) demixing, as well as a solid-solid ($S1$-$S2$) phase separation. In this case, the three phases coexist at a triple point. 

It is interesting to remark that the cubic Husimi lattice solution of the 0NN-2NN mixture presents a fluid-fluid demixing \cite{Nathann19}, which is absent in the mixtures analyzed here. We have indeed scanned the parameter space looking for additional phases, even metastable ones, but it seems that only the three ($F$, $S1$ and $S2$) phases discussed above exist in our approach. These results - together with the ones for the 0NN-1NN model on the square lattice, where no fluid-fluid demixing was observed \cite{poland,Jim01,tiago15} - indicate that the condition for observation of fluid-fluid transitions in such systems might be $k$NN-$k'$NN with $k' \geq k+2$ or, maybe, it is limited to 0NN-$k$NN mixtures, with $k \geq 2$. This is also an important issue to be addressed in future works.

Finally, let us remark that a density anomaly 
is present in the 0NN-1NN mixture, but it is absent in the 1NN-2NN one. Since this anomaly, which might be important for understanding more complex fluids, have already been observed in previous studies of the 0NN-1NN mixture \cite{tiago11,tiago15} and also in the 0NN-2NN model \cite{Nathann19}, its absence in the 1NN-2NN system suggests that the presence of point (0NN) particles in the mixtures is imperative for its existence. In fact, in $0$NN-$k$NN mixtures, only the fluid phase is expected to be stable in the phase diagram for small enough activities ($z_k$) of the larger $k$NN particles, so that it shall exist for $z_0$ ranging from $0$ to $\infty$. This was indeed observed in all mixtures of this type investigated so far. For mixtures of the type $k$NN-$k'$NN, with $0 < k < k'$, the fluid phase is expected to be stable only in a limited region of parameter space (for small $z_k$ and $z_{k'}$), since it shall present transitions (at least) for the two solid phases associated with the ordering of both $k$NN and $k'$NN particles, as indeed observed here in the 1NN-2NN model. In this scenario, it may be the case that the density of small particles $\rho_k$ does not become so large within the fluid phase to allow the onset of the anomalous behavior, which certainly explain its absence in 1NN-2NN. For other mixtures of this type, however, there is no guarantee that this shall happen. So, once again, more studies of these $k$NN mixtures are important to unveil what are the conditions for the appearance of such entropy-driven anomaly. We notice that the symmetries of the ordered phases of $k$NN systems with $k>2$ cannot be captured by a Husimi lattice built with cubes, so that other methods, such as Monte Carlo simulations, shall be employed to answer the important questions raised here.

\acknowledgments

This work is partially supported by CNPq, CAPES and FAPEMIG (Brazilian agencies).

\appendix
\label{apRRs}
\section{Additional details on the solutions of the models on the Husimi lattice}

\subsection{Recursion relations for the 0NN-1NN mixture}

The recursion relations (RRs) for the partial partition functions (ppf's) of the 0NN-1NN mixture, defined and obtained as explained in Sec. \ref{secModel}, when the root site is in the sublattice $A$, are given by

\begin{widetext}
\begin{subequations}
\begin{eqnarray}
 a'_{\varnothing}&=&a_\varnothing^3 b_\varnothing^4\left[3 z_1^2 A_{1}^2 + z_1^3 A_{1}^3 + z_0^3 A_{0}^3 (1 + z_0 B_{0})^4 + 3 z_0^2 A_{0}^2 (1 + z_0 B_{0}) (z_1 A_{1} + (1 + z_0 B_{0})^3) \right. \\ \nonumber
 && \left.+ 3 z_1 A_{1} (1 + z_0 B_{0} + z_1 B_{1}) + (1 + z_0 B_{0} + z_1 B_{1})^4 + 3 z_0 A_{0} (z_1^2 A_{1}^2 + 2 z_1 A_{1} (1 + z_0 B_{0}) + (1 + z_0 B_{0})^3 (1 + z_0 B_{0} + z_1 B_{1}))\right]
\end{eqnarray}
\begin{eqnarray}
 a'_{0}&=&a_\varnothing^3 b_\varnothing^4 \left[3 z_1^2 A_{1}^2 + z_1^3 A_{1}^3 + 3 z_1 A_{1} (1 + z_0 B_{0}) + z_0^3 A_{0}^3 (1 + z_0 B_{0})^4 + 3 z_0^2 A_{0}^2 (1 + z_0 B_{0}) (z_1 A_{1} + (1 + z_0 B_{0})^3) \right. \\ \nonumber
 &&\left.+ 3 z_0 A_{0} (z_1^2 A_{1}^2 + 2 z_1 A_{1} (1 + z_0 B_{0}) + (1 + z_0 B_{0})^4) + (1 + z_0 B_{0})^3 (1 + z_0 B_{0} + z_1 B_{1})\right]
\end{eqnarray}
\begin{eqnarray}
 a'_{1}&=&a_\varnothing^3 b_\varnothing^4 \left[1 + 3 z_1 A_{1} + 3 z_1^2 A_{1}^2 + z_1^3 A_{1}^3 + z_0 B_{0} + z_0^3 A_{0}^3 (1 + z_0 B_{0}) + 3 z_0^2 A_{0}^2 (1 + z_1 A_{1} + z_0 B_{0}) \right. \\ \nonumber
 && \left. + 3 z_0 A_{0} (1 + 2 z_1 A_{1} + z_1^2 A_{1}^2 + z_0 B_{0}) + z_1 B_{1} \right.]
\end{eqnarray}
\label{RRs0NN1NN}
\end{subequations}
\end{widetext}
where the capital letters $A_j$ and $B_j$ denote the ratios $A_j=a_j/a_\varnothing$ and $B_j=b_j/b_\varnothing$, for $j=0,1$. The RRs for the sublattice $B$ are given by a simple permutation between $a$ ($A$) and $b$ ($B$) in the expressions above. Moreover, the RRs for the ratios are obtained by dividing the equations \ref{RRs0NN1NN}, namely, $A'_{j} = a'_{j}/a'_{\varnothing}$ and $B'_{j} = b'_{j}/b'_{\varnothing}$, with $j=0,1$.

\subsection{Solution of the 1NN-2NN mixture}

For the 1NN-2NN mixture, the RRs for the ppf's, when the root site is in sublattice $A_1$, are given by

\begin{widetext}
\begin{subequations}
\begin{eqnarray}
 a'_{1,\varnothing}&=& b_{1,\varnothing} c_{1,\varnothing} d_{1,\varnothing} a_{2,\varnothing} b_{2,\varnothing} c_{2,\varnothing} d_{2,\varnothing}\left[1+z_1 (C_{1,1}+B_{2,1}+A_{2,1}+D_{1,1}+C_{2,1}+B_{1,1}+z_1 (A_{2,1} C_{2,1}+D_{1,1} B_{1,1}\right. \\ \nonumber
 &&\left. +B_{2,1} (A_{2,1}+C_{2,1}+B_{1,1}+ z_1 A_{2,1} C_{2,1})+C_{1,1} (D_{1,1}+C_{2,1}+B_{1,1}+ z_1 D_{1,1} B_{1,1}))\right. \\ \nonumber
 &&\left. +D_{2,1} (1+z_1 (A_{2,1}+D_{1,1}+C_{2,1}+ z_1 A_{2,1} C_{2,1} +B_{2,1} (1+ z_1 A_{2,1}) (1+ z_1 C_{2,1}))))\right. \\ \nonumber
 &&\left. +z_2 (D_{2,2}+C_{1,2}+B_{2,2}+A_{2,2}+D_{1,2}+C_{2,2}+B_{1,2})+ z_1 z_2 (D_{1,1} D_{2,2} +D_{2,1} D_{1,2}\right. \\ \nonumber
 &&\left. +C_{1,1} C_{2,2}+C_{2,1} C_{1,2} +B_{1,1} B_{2,2} +B_{2,1} B_{1,2}) + z_2^2 (D_{1,2} D_{2,2} + C_{1,2} C_{2,2} + B_{1,2} B_{2,2})\right]
\end{eqnarray}
\begin{eqnarray}
 a'_{1,1}&=& b_{1,\varnothing} c_{1,\varnothing} d_{1,\varnothing} a_{2,\varnothing} b_{2,\varnothing} c_{2,\varnothing} d_{2,\varnothing}\left[1 + z_1 (A_{2,1} +  D_{1,1} + B_{1,1} + z_1 D_{1,1} B_{1,1}) + z_2 A_{2,2} \right. \\ \nonumber
 &&\left. + z_1 C_{1,1} (1 + z_1 D_{1,1}) (1 + z_1 B_{1,1})\right]
\end{eqnarray}
\begin{eqnarray}
 a'_{1,2}&=& b_{1,\varnothing} c_{1,\varnothing} d_{1,\varnothing} a_{2,\varnothing} b_{2,\varnothing} c_{2,\varnothing} d_{2,\varnothing}\left(1 + z_1 A_{2,1} + z_2 A_{2,2}\right)
\end{eqnarray}
\label{RRs1NN2NN}
\end{subequations}
\end{widetext}
where $A_{i,j}$, $B_{i,j}$, $C_{i,j}$ and $D_{i,j}$, with $i=1,2$ and $j=1,2$, are the ratios defined by:
\begin{equation}
 A_{i,j}=\frac{a_{i,j}}{a_{i,\varnothing}},\phantom{...}B_{i,j}=\frac{b_{i,j}}{b_{i,\varnothing}},\phantom{...}C_{i,j}=\frac{c_{i,j}}{c_{i,\varnothing}},\phantom{...}D_{i,j}=\frac{d_{i,j}}{d_{i,\varnothing}}.\nonumber
\end{equation}
The RRs for such ratios are defined similarly, by dividing the expressions \ref{RRs1NN2NN} with index $j \neq \varnothing$ by the ones with $j=\varnothing$, for a given sublattice. 
From cyclic permutations of the sublattice labels: $A_1 \rightarrow B_2$, $B_2 \rightarrow C_1$, $C_1 \rightarrow D_2$, $D_2 \rightarrow A_1$, together with $C_2 \rightarrow D_1$, $D_1 \rightarrow A_2$, $A_2 \rightarrow B_1$, $B_1 \rightarrow C_2$, in this order (see Fig. \ref{fig1}e), we can obtain the RRs for the ppf's and ratios for the sublattices $D_2$, $C_1$ and $B_2$. Then, the RRs for the sublattices $A_2$, $B_1$, $C_2$ and $D_1$ can be obtained from the ones for $A_1$, $B_2$, $C_1$ and $D_2$, respectively, by exchanging all the sublattice indexes $i$ ($1 \leftrightarrow 2$). 

The partition function of the 1NN-2NN mixture reads
\begin{eqnarray}
 Y&=&a_{1,\varnothing}a'_{1,\varnothing} + z_1 a_{1,1}a'_{1,1} + z_2 a_{1,2}a'_{1,2}\\ \nonumber
  &=&  a_{1,\varnothing} b_{1,\varnothing} c_{1,\varnothing} d_{1,\varnothing} a_{2,\varnothing} b_{2,\varnothing} c_{2,\varnothing} d_{2,\varnothing} y,
\end{eqnarray}
where $y$ depends only on the ratios at the fixed point and on the activities ($z_1$, $z_2$). From the expanded expression for $Y$, we can calculate the densities of small and large particles in a given sublattice, at the central cube, using definitions analogous to those in Eq. \ref{eqDensities}.

The bulk free energy (per site) for the 1NN-2NN mixture is given by
\begin{equation}
 \phi_b = -\frac{1}{8} \ln \left[ \frac{A_{1,\varnothing} B_{1,\varnothing} C_{1,\varnothing} D_{1,\varnothing} A_{2,\varnothing} B_{2,\varnothing} C_{2,\varnothing} D_{2,\varnothing}}{y^{6}} \right],
\end{equation}
where 
\begin{equation}
 A_{1,\varnothing}\equiv\frac{a'_{1,\varnothing}}{b_{1,\varnothing} c_{1,\varnothing} d_{1,\varnothing} a_{2,\varnothing} b_{2,\varnothing} c_{2,\varnothing} d_{2,\varnothing}},
\end{equation}
and the other ones can be obtained from this expression by using the same scheme of permutations described above for determining the ratios for other sublattices.

\end{document}